# MAXIMUM LIKELIHOOD ESTIMATION OF CLOUD HEIGHT FROM MULTI-ANGLE SATELLITE IMAGERY

By E. Anderes[1], B. Yu[2], V. Jovanovic, C. Moroney, M. Garay, A. Braverman and E. Clothiaux

*University of California at Davis, University of California at Berkeley, California Institute of Technology, California Institute of Technology, Raytheon Corporation, California Institute of Technology and Pennsylvania State University*

We develop a new estimation technique for recovering depth-of-field from multiple stereo images. Depth-of-field is estimated by determining the shift in image location resulting from different camera viewpoints. When this shift is not divisible by pixel width, the multiple stereo images can be combined to form a super-resolution image. By modeling this super-resolution image as a realization of a random field, one can view the recovery of depth as a likelihood estimation problem. We apply these modeling techniques to the recovery of cloud height from multiple viewing angles provided by the MISR instrument on the Terra Satellite. Our efforts are focused on a two layer cloud ensemble where both layers are relatively planar, the bottom layer is optically thick and textured, and the top layer is optically thin. Our results demonstrate that with relative ease, we get comparable estimates to the M2 stereo matcher which is the same algorithm used in the current MISR standard product (details can be found in [*IEEE Transactions on Geoscience and Remote Sensing* **40** (2002) 1547–1559]). Moreover, our techniques provide the possibility of modeling all of the MISR data in a unified way for cloud height estimation. Research is underway to extend this framework for fast, quality global estimates of cloud height.

**1. Introduction.** The motivation for this paper comes from the problem of recovering cloud height from the Multi-Angle Imaging SpectroRadiometer

Received May 2008; revised March 2009.

[1]Supported in part by an NSF Postdoctoral Fellowship DMS-05-03227 and JPL Grant 1302389.

[2]Supported in part by Grants NSF DMS-06-05165, ARO W911NF-05-1-0104, NSFC-60628102, a Guggenheim Fellowship in 2006 and a grant from MSRA.

*Key words and phrases.* Random fields, cloud height estimation, super resolution, multiangle imaging spectroradiometer, depth-of-field, maximum likelihood estimation.

(MISR) instrument, launched in December 1999 on the NASA EOS Terra Satellite. Clouds play a major role in determining the Earth's energy budget. As a result, monitoring and characterizing the distribution of clouds becomes important in global studies of climate. The MISR instrument produces images (275 m resolution) in the red band over a swath width of 360 km for nine camera angles: $70°$, $60°$, $45.6°$, $26.1°$ forward angles, a nadir view at $0°$, and aft angles $26.1°$, $45.6°$, $60°$, $70°$ (referred to as Df, Cf, Bf, Af, An, Aa, Ba, Ca and Da respectively). By taking advantage of the image displacement that results from multi-angle image geometry, one can recover cloud top height and cloud motion (where cloud motion is determined from wind). Unfortunately, transparency, multiple layers, occlusion and height discontinuities present challenges for cloud height estimation. In this paper we apply new statistical techniques for estimating cloud height and attempt to recover cloud height for a two layer ensemble: an optically thin top layer over a textured bottom layer.

The image displacement that results from ground registration is, almost always, not divisible by the pixel width. By taking advantage of this offset, one can construct a super-resolution image from the different viewing angles. It is this super-resolution image that we model as a discrete sample from a realization of a continuous Gaussian random field. Under this paradigm, estimating height and wind can be viewed as a statistical likelihood estimation problem. There are several advantages of this approach. First, by changing the model of the latent continuous image, one can change the matching characteristics of the algorithm and, potentially, optimize the matching for different cloud ensembles. Second, the super-resolution framework extends naturally when there are more than two camera angles. Finally, the modeling of the super-resolution image gives a unified way of estimating sub-grid-scale displacement.

There is a considerable amount of existing literature on both the recovery of three dimensional structure from multiple stereo images and constructing super-resolution images from multiple stereo images. The literature on both problems is vast and spans over at least 20 years. We refer readers to two reviews, Brown, Burschka and Hager (2003) and Dhond and Aggarwal (1989), on the recovery of depth-of-field and two reviews, Farsiu et al. (2004) and Park, Park and Kang (2003), on the super-resolution problem. To the authors' knowledge, the two problems have yet to be considered concurrently for recovery of depth-of-field which is the focus of the current paper.

The rest of the paper is organized as follows. In Section 2 we describe our new technique for estimating depth-of-field from multiple images taken from different viewpoints. In Section 3 we show how to use this super-resolution framework for estimating cloud height from the MISR data. Finally, Section 4 presents our test results for cloud height estimation.

**2. The super-resolution likelihood.** In this section we describe our technique for recovering depth-of-field from multiple stereo images. We start with our notation for nonuniformly sampled images and finish with a presentation of our new estimation methodology that uses super-resolution techniques and random field models to define a likelihood for estimating depth-of-field. This exposition is done as general as possible to avoid letting the details of the cloud height problem obstruct the general estimation procedure (the details of the cloud height estimation are then presented in Section 3).

2.1. *Image notation.* Even though it is easy to visualize the construction of a super-resolution image, the mathematical notation to express this construction is somewhat clumsy. The basic object for our notation is an image, denoted $(\mathbf{x}, \mathbf{y})$. The pixel locations are encoded in the ordered list of spatial coordinates $\mathbf{x}$ and the corresponding gray values, or radiances, are encoded in the vector $\mathbf{y}$. In the regression setting, one might consider $\mathbf{x}$ to be a matrix with two columns where each row is a pixel location. However, we prefer to use the 'list' characterization so that, for example, a function $f : \mathbb{R}^2 \to \mathbb{R}$ evaluated at a list of locations $\mathbf{x}$, denoted $f(\mathbf{x})$, will represent component-wise evaluation of $f$ on each element of the list (rather than row-wise evaluation using the regression notation). In this way, shifting all the spatial locations in $\mathbf{x}$ by the same $\delta \in \mathbb{R}^2$ can be written $\mathbf{x} + \delta$.

The pixel locations $\mathbf{x}$ allow us to define nonuniformly sampled images. Therefore, when working with one image, $\mathbf{x}$ serves to specify the overall structure of the image. When working with multiple images, the individual pixel locations may be useful for comparing locations across multiple images. For example, if one has $n$ images on a square grid of equal size, it may make sense to set the lower left pixel at the origin, all with the same pixel spacing. Multiple images will be written with superscripts $(\mathbf{x}^{(1)}, \mathbf{y}^{(1)}), \ldots, (\mathbf{x}^{(n)}, \mathbf{y}^{(n)})$, reserving subscripts to denote the elements of a list. In particular, $\mathbf{x}_j$ and $\mathbf{y}_j$ denotes the $j$th pixel and the corresponding $j$th gray value or radiance. For example, a $2 \times 2$ gray level image could be written $\mathbf{x} = ((0,0), (0,1), (1,0), (1,1))$, $\mathbf{y} = (0.4, 0.2, 0.1, 0.7)^T$ so that the second pixel has coordinates $\mathbf{x}_2 = (0,1)$ with gray value $\mathbf{y}_2 = 0.2$.

Now the super-resolution image is defined by direct concatenation of the pixel locations and the gray values. In particular, let $(\mathbf{x}^{(1)}, \mathbf{y}^{(1)})$ and $(\mathbf{x}^{(2)}, \mathbf{y}^{(2)})$ be two images which we want to overlay to construct a super-resolution image. The intuition is that these two images are of the same physical object but projected on different pixel grids (see Figure 1). The super-resolution image is defined as $\left( \binom{\mathbf{x}^{(1)}}{\mathbf{x}^{(2)}}, \binom{\mathbf{y}^{(1)}}{\mathbf{y}^{(2)}} \right)$. It may be the situation that the first image $(\mathbf{x}^{(1)}, \mathbf{y}^{(1)})$ needs to be shifted by some vector $\delta \in \mathbb{R}^2$ before it is overlaid with the second image $(\mathbf{x}^{(2)}, \mathbf{y}^{(2)})$. The process of shifting the first image by $\delta$ then overlaying the two to construct a super-resolution image is written $\left( \binom{\mathbf{x}^{(1)}+\delta}{\mathbf{x}^{(2)}}, \binom{\mathbf{y}^{(1)}}{\mathbf{y}^{(2)}} \right)$.

**Image frame from the multiple cameras**

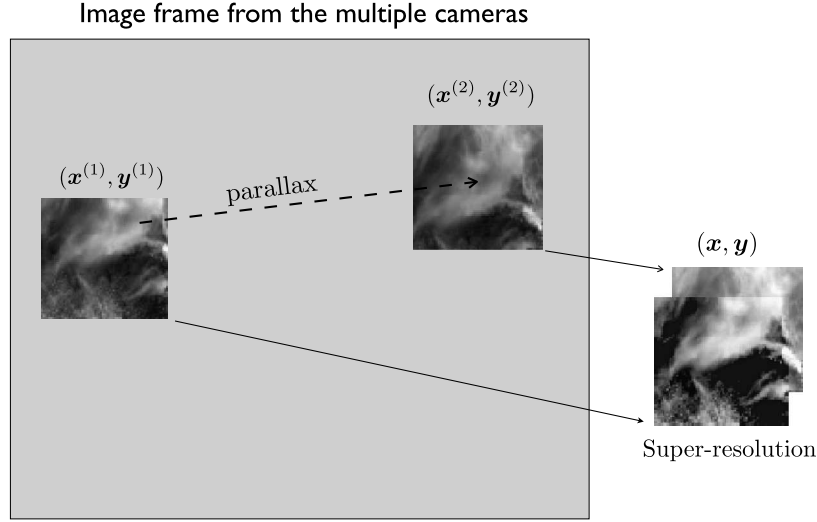

FIG. 1. *An illustration of the construction of the super-resolution image* $(\mathbf{x}, \mathbf{y})$. *The object captured by* $(\mathbf{x}^{(1)}, \mathbf{y}^{(1)})$ *shifts to* $(\mathbf{x}^{(2)}, \mathbf{y}^{(2)})$ *in the next camera angle. If the parallax is known and is not a multiple of the grid spacing, the two image patches can be overlaid to create a super-resolution image.*

2.2. *Super-resolution, random fields and depth-of-field.* We start with $n$ images, each representing different pictures of the same object taken from $n$ different camera viewpoints (see the first illustration of Figure 2). Since the cameras are from different viewpoints, the object will appear in different image locations in each camera (see the second illustration of Figure 2). We define *parallax* as the difference in image location of the object in each camera. Notice that once the geometry of the camera configuration is fixed, the parallax is completely characterized by the distance, $d$, of the object to one of the cameras. Typical estimation of $d$ (i.e., depth-of-field) amounts to estimating the parallax of the object between each pair of cameras then using these parallaxes to solve for $d$. Our method, on the other hand, estimates $d$ directly, treating it as an unknown statistical parameter. The advantage is that $d$ completely characterizes the parallax observed between each pair of cameras, thereby reducing the problem to estimating a single parameter and modeling the observations jointly rather than pairwise.

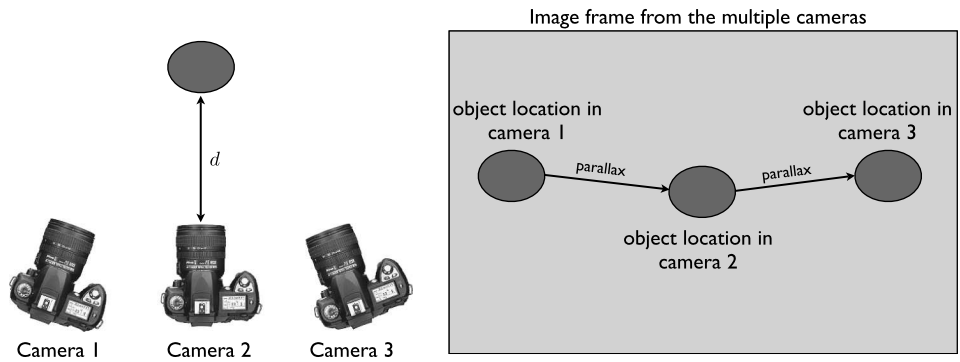

FIG. 2. *An illustration of the distance parameter $d$ and parallax. Once the geometry of the camera array is fixed, the parameter $d$ completely determines the parallax.*

In more detail, consider a physical object captured in the image patch $(\mathbf{x}^{(1)}, \mathbf{y}^{(1)})$ and let $d \in (0, \infty)$ denote the distance of this object to the first camera. As one varies $d$, there will exist different image patches $(\mathbf{x}^{(2)}, \mathbf{y}^{(2)}), \ldots,$ $(\mathbf{x}^{(n)}, \mathbf{y}^{(n)})$ from the subsequent camera angles, all of which capture the object appearing in $(\mathbf{x}^{(1)}, \mathbf{y}^{(1)})$. This implies the existence of shifts $\delta_2, \ldots, \delta_n$ so that the new super-resolution image, $(\mathbf{x}, \mathbf{y})$, given by

$$\text{(1)} \qquad \mathbf{x} := \begin{pmatrix} \mathbf{x}^{(1)} \\ \mathbf{x}^{(2)} + \delta_2 \\ \vdots \\ \mathbf{x}^{(n)} + \delta_n \end{pmatrix}, \qquad \mathbf{y} := \begin{pmatrix} \mathbf{y}^{(1)} \\ \vdots \\ \mathbf{y}^{(n)} \end{pmatrix},$$

is a picture of the same object captured by $(\mathbf{x}^{(1)}, \mathbf{y}^{(1)})$ (see Figure 1). Note that the shifts $\delta_2, \ldots, \delta_n$ and the image patches $(\mathbf{x}^{(2)}, \mathbf{y}^{(2)}), \ldots, (\mathbf{x}^{(n)}, \mathbf{y}^{(n)})$ depend solely on the distance parameter $d$. We call the process of combining image patches to construct a super-resolution image "interlacing."

To compute a likelihood for estimating the parameter $d$, we start by supposing there exists a latent continuous function $Y : \mathbb{R}^2 \to \mathbb{R}$ that models the continuous image in the local patch $(\mathbf{x}^{(1)}, \mathbf{y}^{(1)})$. In other words, $Y(\mathbf{x}^{(1)}) = \mathbf{y}^{(1)}$. Note that $Y(\mathbf{x})$ denotes the list obtained by evaluating $Y$ at each pixel location in the list $\mathbf{x}$. For the true distance parameter $d$, the super-resolution image $(\mathbf{x}, \mathbf{y})$ will also satisfy $Y(\mathbf{x}) = \mathbf{y}$. The wrong distance parameter will interlace images of different regions which will result in a 'noisy' super-resolution image for which it will be difficult to find a continuous interpolator that satisfies $Y(\mathbf{x}) = \mathbf{y}$. By putting a probability measure $\mathbb{P}$ on the latent continuous image $Y$, we can estimate the distance of the object by minimizing the following negative log likelihood:

$$\text{(2)} \qquad -\ell(d) := -\log \mathbb{P}(Y(\mathbf{x}) = \mathbf{y}),$$

where the super-resolution image $(\mathbf{x}, \mathbf{y})$, depending on $d$, is constructed as in (1).

An equivalent way to specify the log-likelihood (2) is to simply claim that there exists image patches $(\mathbf{x}^{(2)}, \mathbf{y}^{(2)}), \ldots, (\mathbf{x}^{(n)}, \mathbf{y}^{(n)})$ and location shifts $\delta_2, \ldots, \delta_n$, all depending on the parameter $d$, such that $Y(\mathbf{x}^{(1)}) = \mathbf{y}^{(1)}$ and

$$\text{(3)} \qquad \mathbf{y}^{(k)} = Y(\mathbf{x}^{(k)} + \delta_k), \qquad k = 2, \ldots, n.$$

Since $Y$ does not depend on $k$ in the above equation, the super-resolution image defined by (1) satisfies $\mathbf{y} = Y(\mathbf{x})$. Therefore, under the model $\mathbb{P}$, the joint distribution of $\mathbf{y}$ is the same as that obtained in (2). This alternative way of expressing the super-resolution likelihood will become useful in the following sections for introducing nuisance parameters associated with the MISR cameras.

**3. Cloud height estimation.** This section contains the modeling details for cloud height estimation from MISR satellite data. In our view, one of the advantages of the estimation techniques outlined in the previous section is the ease in which the technical details of a particular observation scenario can be incorporated into the recovery of depth-of-field. We demonstrate this flexibility in our application of this methodology to cloud height estimation. This section starts with a discussion of the relationship between parallax, cloud height and wind. Section 3.1 presents a summary of all the modeling assumptions used to compute the maximum likelihood estimation of height. A more thorough explanation of the modeling assumptions and the techniques for computing the likelihood can be found in Sections 3.1.1, 3.1.2 and 3.1.3.

For the MISR data, instead of estimating distance to the satellite, we want to estimate cloud height and wind. Since the parallax of a cloud patch is completely determined by height and wind, the techniques from Section 2 are easily applicable. Indeed, one of the main features of the methodology developed in the last section is that it relates all the observed parallaxes to a single parameter $d$. In the MISR data, the parameter $d$, that is, distance of the object to the satellite, is directly related to cloud height (since the height of a cloud determines the distance to the satellite). However, there is an added complication of cloud movement as a function of wind. Since wind will also effect the image location of a cloud in each camera, we include it as a parameter so that now height $h$ and a wind velocity $\mathbf{v} := (v_1, v_2)$ completely determine parallax and the construction of the super-resolution image $(\mathbf{x}, \mathbf{y})$.

For the MISR images there is an approximate linear relationship that relates wind, cloud height and image parallax. Let $t^{(k)}$ denote the fly-over time delay (in seconds) for the $k$th camera and $x^{(k)}$ the along-track image location of a local cloud region in camera image $k$. When $h$ is in meters, $\mathbf{v}$ is in meters per second, and the pixel locations specified by $\dot{\mathbf{x}}^{(k)}$ all represent the same 275 meter grid, the linear equation that relates wind, height and along-track parallax are

$$(4) \qquad v_2(t^{(i)} - t^{(j)}) + h(\tan(\theta^{(i)}) - \tan(\theta^{(j)})) = x^{(i)} - x^{(j)},$$

where $\theta^{(k)}$ is the angle of each camera. The across-track parallax is given by $v_1(t^{(i)} - t^{(j)})$. For more details see Diner et al. (1997).

Before we continue, we mention that since clouds behave dynamically there is the possibility of a change in cloud structure within the time delay of the different cameras. However, since this delay is at most 7 minutes (the approximate delay from the Df camera to the Da camera) and the pixel patches we will be using correspond to about 4 square kilometers (15-by-15 pixel patches), we will tacitly assume this effect is negligible.

3.1. *Detailed summary of the model.* A major obstacle in using the methods in Section 2 for the MISR data is that the images from different angles often have different overall brightness. Therefore, it becomes difficult to directly interlace patches from different images. In what follows we model this brightness change as a linear correction for each camera.

Suppose the image patch $(\mathbf{x}^{(1)}, \mathbf{y}^{(1)})$ captures a small cloud region and let $Y$ be the continuous representation of this image so that $Y(\mathbf{x}^{(1)}) = \mathbf{y}^{(1)}$. The true height and wind, $(h, \mathbf{v})$, allow us to retrieve patches $(\mathbf{x}^{(2)}, \mathbf{y}^{(2)}), \ldots,$ $(\mathbf{x}^{(n)}, \mathbf{y}^{(n)})$ from the subsequent camera angles, each capturing the same cloud region projected on potentially shifted grids. For the remainder of the paper, $n$ denotes the number of cameras and $m$ denotes the number of pixels in each patch $(\mathbf{x}^{(k)}, \mathbf{y}^{(k)})$, $k = 1, \ldots, n$. Our model for the gray values $\mathbf{y}^{(k)}$ is summarized by

$$(5) \qquad \mathbf{y}^{(k)} = \sigma_k Y(\mathbf{x}^{(k)} + \delta_k) + A_k \mathbf{x}^{(k)} + b_k$$

for cameras $k = 2, \ldots, n$, where

- $Y$ is the latent continuous cloud image, modeled by a Gaussian random field with Matérn covariance function $\mathcal{K}$ with parameters $(\sigma, \rho, \nu) = (1, 4, 4/3)$ (see Section 3.1.1) so that

$$\mathrm{cov}(Y(\mathbf{t}), Y(\mathbf{s})) = \mathcal{K}(|\mathbf{t} - \mathbf{s}|),$$

  for all $\mathbf{t}, \mathbf{s} \in \mathbb{R}^2$ where $|\cdot|$ denotes Euclidean distance;
- $\sigma_k$ is a multiplicative brightness correction for each camera;
- $\delta_k$ is a two dimensional shift vector which is completely determined by $(h, \mathbf{v})$ and plays the role of shifting the pixel locations $\mathbf{x}^{(1)}, \ldots, \mathbf{x}^{(n)}$ so that they are interlaced;
- $A_k$ is a $2 \times 1$ matrix which represents an affine brightness correction for each patch;
- $b_k$ is an overall constant additive correction.

Notice that $Y$ models the latent cloud image for each camera so we obtain a joint model for the full super-resolution vector $\mathbf{y} := (\mathbf{y}^{(1)\,T}, \ldots, \mathbf{y}^{(n)\,T})^T$. By the Gaussian assumption on $Y$, we have

$$(6) \qquad \mathbf{y}^{(k)} \sim \mathcal{N}(A_k \mathbf{x}^{(k)} + b_k, \sigma_k^2 \Sigma_k),$$

where $\Sigma_k := (\mathcal{K}(|\mathbf{x}_i^{(k)} - \mathbf{x}_j^{(k)}|))_{ij}$.

Since we have introduced an additive and multiplicative brightness correction on each patch, it is no longer true that the super-resolution image satisfies $\mathbf{y} = Y(\mathbf{x})$, where $\mathbf{x}$ is defined by (1). What is true, however, is that $\mathbf{y} = \mathrm{diag}(\sigma_1 I_m, \ldots, \sigma_n I_m) Y(\mathbf{x}) + \mu$, where $\mu$ is a vector with $k$th block $A_k \mathbf{x}^{(k)} + b_k$ and $I_m$ is the $m \times m$ identity matrix. The main difficulty for estimation of $(h, \mathbf{v})$ now becomes devising techniques for dispensing with the nuisance parameters $\sigma_k, A_k, b_k$. This is the main contribution of Sections 3.1.2 and 3.1.3. Both sections use REML techniques [see Stein (1999)] to filter out the additive brightness corrections $A_k \mathbf{x}^{(k)} + b_k$. Dealing with the multiplicative nuisance parameters $\sigma_k$ is more difficult and requires separate treatments for high and low clouds. For high clouds, a brightness stabilization allows us to marginalize out $\sigma_k$. For low clouds, there is no closed form solution for the marginalized likelihood and we resort to an approximate profile likelihood.

REMARK.    The affine correction $A_k \mathbf{x}^{(k)} + b_k$, rather than a higher order polynomial, was chosen for two reasons. First, the size of the pixel patch is relatively small. Small enough, in fact, so that one might expect a linear Taylor approximation to be valid. Second, a linear polynomial seems the natural choice for a once, but not twice, differentiable cloud process, which is derived in the following subsection.

3.1.1. *Gaussian random field model for* $\mathbb{P}$. In (5) we use a Matérn autocovariance function $\mathcal{K}$ [with parameters $(\sigma, \rho, \nu) = (1, 4, 4/3)$] to model the covariance structure of the latent continuous cloud image $Y$. This section outlines our motivation for our choice of the Matérn autocovariance function and the specific parameter values. The basic idea is to use the observed fractal behavior of clouds [see Cahalan and Snider (1989), Barker and Davies (1992), Várnai (2000), Oreopoulos et al. (2000) and Davis et al. (1997) from the atmospheric science literature] to model the fractal nature of $Y$ after adjusting for pixel averaging. Indeed, one of the advantages of the the Matérn autocovariance function is the flexibility it provides for modeling the fractal behavior of $Y$.

The Matérn autocovariance function $\mathcal{K}$ is defined by

$$\mathcal{K}(r) = \frac{\sigma}{2^{\nu-1}\Gamma(\nu)} \left( \frac{2\nu^{1/2} r}{\rho} \right)^{\nu} \mathcal{K}_{\nu}\left( \frac{2\nu^{1/2} r}{\rho} \right),$$

where $\mathcal{K}_{\nu}$ is the modified Bessel function [see Stein (1999)]. We model $Y$, the continuous cloud image, as a two dimensional Gaussian random field with covariance structure given by $\mathrm{cov}(Y(s), Y(t)) = \mathcal{K}(|s - t|)$ for all $s, t \in \mathbb{R}^2$ with parameters $(\sigma, \rho, \nu) = (1, 4, 4/3)$. The parameter $\sigma$ is the variance $\mathrm{var}[Y(t)]$ at any fixed point $t \in \mathbb{R}^2$. The parameter $\rho$ serves as the range parameter for $Y$ so that $\text{cov}(Y(s), Y(t))$ quickly becomes negligible when the lag $|s - t|$ is larger than $\rho$. The $\nu$ parameter controls the smoothness of the process.

In our analysis the range parameter $\rho$ is set to 1100 meters (4 pixel widths) and the variance parameter $\sigma$ is 1. With the exception of either very small or very large $\rho$, different values of $\rho$, $\sigma$ do not severely effect our height estimates. In the case of very large $\rho$ it seems that the local patch size becomes too small, compared to the range, for appropriate modeling. When $\rho$ is very small, the estimation of height and wind becomes ineffective since most of the interlaced images are modeled as nearly uncorrelated.

Here we derive our justification for $\nu = 4/3$ as a plausible value for the smoothness parameter. In our analysis, the vector $\mathbf{y}^{(k)}$ represents the log of registered bidirectional reflectance factor (BRF) values from the $k$th camera. Let $I_{\text{BRF}}$ denote the latent continuous BRF image so that $Y = \log(\varphi * I_{\text{BRF}})$, where the pixelation is represented by the convolution kernel $\varphi = \mathbb{I}_{[-1/2, 1/2]^2}$ for the indicator function $\mathbb{I}$. Let $\mathcal{K}_{\text{BRF}}$ be the autocovariance function for the continuous BRF image $I_{\text{BRF}}$. The two dimensional Fourier transform of $\mathcal{K}_{\text{BRF}}$ gives the spectral density

$$f_{I_{\text{BRF}}}(\omega) := \frac{1}{(2\pi)^2} \int_{\mathbb{R}^2} \exp(-ix^T \omega) \mathcal{K}_{\text{BRF}}(x) \, dx,$$

where $x = (x_1, x_2)$ and $\omega = (\omega_1, \omega_2)$ are both two dimensional vectors. There is evidence to suggest that Kolmogorov's 5/3 scaling law holds for $f_{I_{\text{BRF}}}$ [see Cahalan and Snider (1989), Barker and Davies (1992), Várnai (2000), Oreopoulos et al. (2000) and Davis et al. (1997)]. Under the additional assumption of isotropy, the scaling law implies the two dimensional spectral density satisfies $f_{I_{\text{BRF}}}(\omega) \asymp \frac{1}{|\omega|^{5/3+1}}$ as $|\omega| \to \infty$. Therefore, the spectral density of the pixelated process $\varphi * I_{\text{BRF}}$ is given by

$$(7) \qquad f_{\varphi * I_{\text{BRF}}}(\omega) = |\hat{\varphi}(\omega)|^2 f_{I_{\text{BRF}}}(\omega) \asymp \text{sinc}^2\left(\frac{\omega_1}{2}\right) \text{sinc}^2\left(\frac{\omega_2}{2}\right) \frac{1}{|\omega|^{5/3+1}},$$

as $|\omega| \to \infty$. Let $f_Y$ denote the spectral density for the Matérn autocovariance model for $Y$. We believe a plausible value for the parameter $\nu$, for the random field $Y = \log(\varphi * I_{\text{BRF}})$, corresponds to matching the power-law decay of $f_Y$ to that of $f_{\varphi * I_{\text{BRF}}}$, in the coordinate directions. In particular, by (7) and the properties of the Matérn spectral density,

$$f_{\varphi * I_{\text{BRF}}}(\omega) \asymp \frac{\sin^2(\omega_1/2)}{|\omega_1|^{5/3+3}}, \qquad f_Y(\omega) \asymp \frac{1}{|\omega_1|^{2\nu+2}},$$

fixing $\omega_2$ and letting $\omega_1 \to \infty$. Therefore, to match the decay, we set the smoothness parameter $\nu$ to $(5/3 + 1)/2 = 4/3$.

3.1.2. *Low-cloud likelihood.* Here we give some of the computational techniques for dealing with the nuisance parameters when estimating the height of low textured clouds. We start by constructing a matrix $L$ ('$L$' for low) which filters out the dependence of the observations on $A_k$ and $b_k$. Then we construct an approximate profile likelihood to handle the multiplicative nuisance parameters $\sigma_k$.

For low textured clouds we begin by constructing $n$ matrices $L_1, \ldots, L_n$ which annihilate monomials of order at most 1 so that $L_k \mathbf{y}^{(k)} \sim \mathcal{N}(0, \sigma_k^2 \Sigma_k)$. In particular, $L_k$ is a $(m-3) \times m$ matrix with rows composed of linearly independent vectors in the kernel space $\{v : [\phi_0(\mathbf{x}^{(k)}) \ \phi_1(\mathbf{x}^{(k)}) \ \phi_2(\mathbf{x}^{(k)})]^T v^T = \mathbf{0}\}$, where $\phi_0, \phi_1$ and $\phi_2$ are the the the monomials of order at most 1. Now define the matrix $L = \text{diag}(L_1, \ldots, L_n)$ so that $L\mathbf{y} \sim \mathcal{N}(0, \Delta_\sigma \tilde{\Sigma} \Delta_\sigma)$, where $\tilde{\Sigma} := L \Sigma L^T$, $\Sigma := (\mathcal{K}(|\mathbf{x}_i - \mathbf{x}_j|))_{ij}$ and $\Delta_\sigma := \text{diag}(\sigma_1 I_{m-3}, \ldots, \sigma_n I_{m-3})$. Therefore, the likelihood of the observation vector $L\mathbf{y}$, as it depends on $(h, \mathbf{v})$ and $\boldsymbol{\sigma} := (\sigma_1, \ldots, \sigma_n)^T$, can be written

$$\mathcal{L}(h, \mathbf{v}, \boldsymbol{\sigma}|L\mathbf{y}) = \frac{1}{|2\pi\tilde{\Sigma}|^{1/2}} \frac{1}{|\Delta_\sigma|} \exp\left[-\frac{1}{2}(L\mathbf{y})^T \Delta_\sigma^{-1} \tilde{\Sigma}^{-1} \Delta_\sigma^{-1}(L\mathbf{y})\right].$$

At this point there are a few options to dispense with the dependence of the above likelihood on the nuisance parameters $\boldsymbol{\sigma}$. The most desirable option would be to marginalize out $\boldsymbol{\sigma}$ by integrating $\int_{\mathbb{R}_+^n} \mathcal{L}(h, \mathbf{v}, \boldsymbol{\sigma}|L\mathbf{y}) \, d\boldsymbol{\sigma}$. For even moderate block size $m$ and $n = 2$, this becomes computationally formidable. The second option would be to maximize a profile likelihood $\mathcal{L}(h, \mathbf{v}|L\mathbf{y}) := \max_{\boldsymbol{\sigma} \in \mathbb{R}_+^n} \{\mathcal{L}(h, \mathbf{v}, \boldsymbol{\sigma}|L\mathbf{y})\}$. This option is somewhat more computationally tractable but still problematic since there is no closed form for the profile likelihood. The easiest option is to estimate $\sigma_k$ on each patch separately, $\hat{\sigma}_k^2 := (L_k \mathbf{y}^{(k)})^T (L_k \Sigma_k L_k^T)^{-1}(L_k \mathbf{y}^{(k)})/m$, then maximizing a plug-in version of the likelihood $\mathcal{L}(h, \mathbf{v}, \hat{\boldsymbol{\sigma}}|L\mathbf{y})$, where $\hat{\boldsymbol{\sigma}} := (\hat{\sigma}_1, \ldots, \hat{\sigma}_n)^T$. Notice, however, that under the correct $(h, \mathbf{v})$ the patches $\mathbf{y}^{(1)}, \ldots, \mathbf{y}^{(n)}$ are highly correlated, and, therefore, information is lost by estimating $\sigma_k$ separately on each patch. In an attempt to alleviate this problem, we explore a compromise between the full profile likelihood and the overly simplistic case of the plug-in likelihood with $\hat{\boldsymbol{\sigma}}$.

Notice that the quadratic term $(L\mathbf{y})^T \Delta_\sigma^{-1} \tilde{\Sigma}^{-1} \Delta_\sigma^{-1}(L\mathbf{y})$ can be written

$$\|\tilde{\Sigma}^{-1/2} \Delta_\sigma^{-1} L\mathbf{y}\|^2 = \left\|\frac{R_1 L_1 \mathbf{y}^{(1)}}{\sigma_1} + \cdots + \frac{R_n L_n \mathbf{y}^{(n)}}{\sigma_n}\right\|^2 = \boldsymbol{\sigma}^{-T} \tilde{R} \boldsymbol{\sigma}^{-1},$$

where the matrices $R_1, \ldots, R_n$ decompose $\tilde{\Sigma}^{-1/2}$ into block form so that $\tilde{\Sigma}^{-1/2} = (R_1, \ldots, R_n)$ and $\tilde{R}$ is the matrix of inner products $(\langle R_i L_i \mathbf{y}^{(i)}, R_j L_j \mathbf{y}^{(j)}\rangle)_{ij}$. Therefore, the log likelihood can be written

$$\ell(h, \mathbf{v}, \boldsymbol{\sigma}|L\mathbf{y}) = c_1 - \frac{1}{2}\log|\tilde{\Sigma}| - (m-3)\sum_{k=1}^n \log \sigma_k - \frac{1}{2}\boldsymbol{\sigma}^{-T} \tilde{R} \boldsymbol{\sigma}^{-1},$$

where $c_1$ is a constant. For a fixed $(h, \mathbf{v})$, maximizing $\ell$ over $\boldsymbol{\sigma}$ is a convex problem whose stationary point is characterized by

$$(8) \qquad \tilde{R}\boldsymbol{\sigma}^{-1} - (m-3)\boldsymbol{\sigma} = 0.$$

This equation defines the profile likelihood but unfortunately has no closed form solution for general $n$. However, there is considerable research on investigating iterative algorithms for solving (8) [see Marshall and Olkin (1968), Khachiyan and Kalantari (1992) and O'Leary (2003), for example]. The problem with iterative algorithms is that the likelihood computation will be done many times while searching through height and wind over sliding blocks. Therefore, including an iterative algorithm in the likelihood calculation presents some computational problems. As a compromise, we define a "one Newton step" estimate of the stationary point of (8) with an initial starting point $\hat{\boldsymbol{\sigma}}$, the maximum likelihood estimate $\sqrt{(L_k\mathbf{y}^{(k)})^T(L_k\Sigma_kL_k^T)^{-1}(L_k\mathbf{y}^{(k)})/m}$ from each camera separately.

The Newton step is constructed as in Khachiyan and Kalantari (1992) and defined for $\boldsymbol{\sigma}^{-1}$ rather than $\boldsymbol{\sigma}$. In particular, define $F(\boldsymbol{\sigma}^{-1}) = \tilde{R}\boldsymbol{\sigma}^{-1} - (m-3)\boldsymbol{\sigma}$ and $\tilde{\Delta}_{\boldsymbol{\sigma}} := \operatorname{diag}(\sigma_1, \ldots, \sigma_n)$. Now $F(\boldsymbol{\sigma}^{-1} + \boldsymbol{\tau}) = F(\boldsymbol{\sigma}^{-1}) + \tilde{R}\boldsymbol{\tau} + (m-3)\tilde{\Delta}_{\boldsymbol{\sigma}}^2\boldsymbol{\tau}$ plus higher order terms in $\boldsymbol{\tau}$. Setting this linear approximation to zero and solving for $\boldsymbol{\tau}$ gives the Newton step $\boldsymbol{\sigma}^{-1} + \boldsymbol{\tau}$. We use the initial starting point $\hat{\boldsymbol{\sigma}}^{-1}$. Notice the Newton step may result in a negative variance, in which case, the original estimate $\hat{\boldsymbol{\sigma}}$ is used instead of the Newton step. In summary, to estimate $(h, \mathbf{v})$, we define the "one Newton step"

$$(\hat{\boldsymbol{\sigma}}_{\text{newton}})^{-1} := \begin{cases} \hat{\boldsymbol{\sigma}}^{-1} + (\tilde{R} + (m-3)\tilde{\Delta}_{\hat{\boldsymbol{\sigma}}}^2)^{-1}((m-3)\tilde{\Delta}_{\hat{\boldsymbol{\sigma}}}^2 - \tilde{R})\hat{\boldsymbol{\sigma}}^{-1}, \\ \qquad \text{if positive;} \\ \hat{\boldsymbol{\sigma}}^{-1}, \qquad \text{otherwise,} \end{cases}$$

and maximize the plug-in log-likelihood

$$\widehat{(h, \mathbf{v})} := \arg\max_{(h,\mathbf{v})} \ell(h, \mathbf{v}, \hat{\boldsymbol{\sigma}}_{\text{newton}} | L\mathbf{y}).$$

3.1.3. *High-cloud likelihood.* Here we give some of the computational techniques for dealing with the nuisance parameters when estimating the height of high thin clouds. Since these heights are notoriously hard to estimate, we attempt a brightness stabilization on the whole cloud image which allows more local information for estimating cloud height. After this stabilization, we construct a matrix $H$ ('$H$' is for high) which filters out the common value of the stabilized nuisance parameters $A_k$ and $b_k$. For the brightness corrections $\sigma_k$, our stabilization allows us to marginalize out the common value of $\sigma_k$ under a uniform prior which is far more desirable than the approximate profile likelihood techniques needed for low clouds.

One of the difficulties of high clouds is that they are frequently optically thin or partially transparent. This makes estimating parallax difficult because the dominant signal is frequently the background rather than the cloud itself. In an attempt to overcome this difficulty, we estimate a linear transformation on $\mathbf{y}^{(k)}$ for each camera $k = 1, \ldots, n$ to stabilize the brightness corrections modeled by $\sigma_k, A_k$ and $b_k$. A more in-depth discussion on this issue is presented in Section 4. Once stabilization is achieved, we have $\mathbf{y} \sim \mathcal{N}(A\mathbf{x} + b, \sigma^2 \Sigma)$, where $\sigma, A$ and $b$ are the common values of $\sigma_k, A_k$ and $b_k$ respectively and $\Sigma := (\mathcal{K}(|\mathbf{x}_i - \mathbf{x}_j|))_{ij}$. This allows us to find a matrix $H$ for which $H\mathbf{y} \sim \mathcal{N}(\mathbf{0}, \sigma H \Sigma H^T)$. $H$ is defined as the matrix with rows composed of linearly independent vectors in the kernel space $\{v : [\phi_0(\mathbf{x}) \, \phi_1(\mathbf{x}) \, \phi_2(\mathbf{x})]^T v^T = \mathbf{0}\}$. Notice that now $H$ is not block diagonal as it was for $L$. As a consequence, $H\mathbf{y}$ is a vector of length $nm - 3$ versus $(m-3)n$ (for the low cloud estimates), where $n$ and $m$ denote the number of cameras and patch size respectively. This provides more information for matching in an attempt at recovering the high-cloud heights.

The likelihood of $h, \mathbf{v}$ and $\sigma$, given the observation vector $H\mathbf{y}$, has the following form:

$$\mathcal{L}(h, \mathbf{v}, \sigma | H\mathbf{y}) = \frac{1}{\sigma^{nm-3} |2\pi \tilde{\Sigma}|^{1/2}} \exp\left[-\frac{\mathbf{y}^T H^T \tilde{\Sigma}^{-1} H\mathbf{y}}{2\sigma^2}\right],$$

where $\tilde{\Sigma} := H \Sigma H^T$. In contrast with the low-cloud likelihood, we can remove the dependence on the unknown parameter $\sigma$ by marginalizing out under a uniform improper prior

$$\mathcal{L}(h, \mathbf{v} | H\mathbf{y}) \propto \int_{\mathbb{R}_+} \mathcal{L}(h, \mathbf{v}, \sigma | H\mathbf{y}) \, d\sigma = \frac{2^{(nm-4)/2}}{2 |2\pi \tilde{\Sigma}|^{1/2}} \frac{\Gamma((nm-4)/2)}{[\mathbf{y}^T H^T \tilde{\Sigma}^{-1} H\mathbf{y}]^{(nm-4)/2}}.$$

The integral is obtained by the change of variables $x = \sigma^2$ and recognizing an un-normalized inverse gamma density. Therefore, the log-likelihood which is maximized over $(h, \mathbf{v})$ is

$$(9) \qquad \ell(h, \mathbf{v} | H\mathbf{y}) = c_2 - \frac{1}{2} \log |\tilde{\Sigma}| - \frac{nm-4}{2} \log[\mathbf{y}^T H^T \tilde{\Sigma}^{-1} H\mathbf{y}],$$

for some constant $c_2$.

REMARK. Instead of marginalizing out the unknown $\sigma$, one could instead maximize a profile log-likelihood as a function of $(h, \mathbf{v})$, where the profiling is over the unknown $\sigma$. The resulting profile log-likelihood is $c_3 - \frac{1}{2} \log |\tilde{\Sigma}| - \frac{nm-3}{2} \log[\mathbf{y}^T H^T \tilde{\Sigma}^{-1} H\mathbf{y}]$ which, it appears, fails to take into account the loss of degrees of freedom associated with the unknown $\sigma$.

3.1.4. *Simulation study.* Before our methodology is applied to the MISR data, we present a short simulation study that provides some evidence that the approximations used in the above methodology are not only valid in some cases but can also provide an improvement over the M2 algorithm. Although our simulations provide favorable evidence for our methodology, we believe the real strength lies in its flexibility to incorporate statistical principles and science to the problem of cloud height retrieval. Before we continue, we stress that this simulation is designed to test the matching algorithm when the probabilistic mechanism that generates the cloud images is known. In general, this will not be the case. Therefore, we consider simulations of this type to be one component of a full investigation of the potential for the super-resolution methodology.

We have three main objectives for our simulations. The first is to show that one can obtain good cloud-height estimates by correctly specifying the high frequency behavior of the spectral density (as we do in Section 3.1.1) even when the the low frequency behavior is mis-specified. Second, we will argue that the joint modeling of the full super-resolution image, rather than pairwise modeling for each camera, gives better estimates. Finally, we provide evidence that our method can outperform the M2 algorithm.

The simulation is designed to mimic the situation where a small cloud patch appears in two larger images of a single cloud scene projected onto different grids. The goal is to then estimate the location of that cloud patch in the two larger images. To this end, we simulated a random field on a narrow two dimensional strip in $[0, 6/500] \times [0, 1]$ with spacing $1/500$ in the $y$ direction (i.e., vertical) and spacing $3/500$ in the $x$ direction. Three images were extracted by sub-sampling every third pixel in the $y$ direction. Therefore, each image is of the same realization but projected on shifted grids. The last strip is then discarded with the exception of a 3 by 4 pixel patch with the lower left corner at $(0, 0.5040)$. The object is then to find the location of the patch, measured along the $y$-axis, in the other two images. To model the situation where there is a lower dimensional parameter [as in (4) for the cloud examples] that reduces the search space of patch locations, we constructed a latent parameter $d \in [0, 1]$ that relates to the two image locations by $y_1 = d$ and $y_2 = 1.9(0.5040) - 0.9d$, where $y_1$ and $y_2$ denote the unknown locations of the patch in the two images. Notice that this sub-manifold contains the true patch locations (i.e., $y_1 = y_2 = 0.5040$ when $d = 0.5040$). Finally, the second strip was multiplied by 10 and the patch (from the third strip) was multiplied by 5 to model an unknown change in brightness across cameras.

The model used in the simulations is a mean zero Gaussian random field with covariance given by

$$\text{cov}(Y(\mathbf{s}), Y(\mathbf{t})) = \sigma^2 |10(\mathbf{s} - \mathbf{t})|^{8/3} + \sum_{0 \le |\boldsymbol{\alpha}| \le 1} c_{\boldsymbol{\alpha}}(\mathbf{s}) \mathbf{t}^{\boldsymbol{\alpha}}$$

$$+ \sum_{0 \leq |\boldsymbol{\alpha}| \leq 1} c_{\boldsymbol{\alpha}}(\mathbf{t}) \mathbf{s}^{\boldsymbol{\alpha}} + \sum_{0 \leq |\boldsymbol{\alpha}|, |\boldsymbol{\beta}| \leq 1} b_{\boldsymbol{\alpha}, \boldsymbol{\beta}} \mathbf{t}^{\boldsymbol{\alpha}} \mathbf{s}^{\boldsymbol{\beta}},$$

where $\boldsymbol{\alpha}$ and $\boldsymbol{\beta}$ are multi-indices (so that $\mathbf{t}^{\boldsymbol{\alpha}}$ and $\mathbf{s}^{\boldsymbol{\beta}}$ are monomials), $c_{\boldsymbol{\alpha}}$ are functions, $b_{\boldsymbol{\alpha}, \boldsymbol{\beta}}$ are constants, and $\sigma = 15$ [see page 250 of Light and Wayne (1998) for the exact construction of $c_{\boldsymbol{\alpha}}$ and $b_{\boldsymbol{\alpha}, \boldsymbol{\beta}}$ or page 256 of Chilès and Delfiner (1999) for a general discussion on the above covariance function]. This model is useful since it has the same high frequency behavior of the Matérn spectral density when $\nu = 4/3$ (which is the model used in the likelihood computation) but a different low frequency behavior. In particular, the random field used in the simulation has a generalized autocovariance proportional to $|\mathbf{t}|^{8/3}$ and thus a generalized spectral density proportional to $|\boldsymbol{\omega}|^{-8/3-2}$ [see Section 2.9 of Stein (1999)]. In contrast, the spectral density of the Matérn autocovariance function used in Section 3.1.1 is proportional to $(|\boldsymbol{\omega}|^2 + 1/3)^{-4/3-1}$ and therefore has the same power decay at infinity but without the singularity at the origin.

The above simulation was repeated 500 times, estimating the parameter $d$ (the true value at $d = 0.5040$) on each realization using 5 different matching metrics. Table 1 reports the results where the rows are ordered from top to bottom by the root mean squared error (RMSE), smallest to largest. The method with the smallest RMSE, called 'full likelihood,' uses the techniques from Section 3.1.2. The second best method uses the same likelihood technique but only pairwise, one for each image. The fact that the full likelihood technique has a smaller RMSE than the pairwise likelihood provides evidence that there is precision gained by fully modeling the joint distribution of the super-resolution image. The next method, 'no Newton update,' is the same as the full likelihood technique but does not update the scale estimates by the Newton step developed in Section 3.1.2. The method 'M2' uses the matching metric in equation (1) of Muller et al. (2002) for the MISR software. The fact that the full likelihood outperforms these techniques provides evidence that by modeling the probabilistic mechanism which generates the images one can improve the height estimates beyond the M2 algorithm. We also reiterate that the full likelihood only correctly specifies the asymptotic decay of the spectral density, and not the low frequency behavior. Therefore, even with this model mis-specification one still gets an improvement over M2. On the other hand, the worst method, 'wrong $\nu$,' uses a Matérn auto-covariance function with $\nu = 2/3$ (the correct value should be $\nu = 4/3$) to model the second order increments of the super-resolution image, and hence incorrectly specifying the high frequency behavior. The fact that using the wrong $\nu$ provides no improvement suggests that it is more important to correctly specify the high order frequency behavior of the spectral density rather than the low-order frequency.

TABLE 1
*The simulation results (500 independent realizations) for estimating the latent parameter $d$ (the true value is $0.5040$) which determines the location of a simulated patch in two images. The left column indexes the different matching methods. The middle column lists the average estimate. The right column lists the root mean square error (RMSE)*

| Method | Average estimate | RMSE |
|---|---|---|
| Full likelihood | 0.50398 | $2.8460 \times 10^{-4}$ |
| Pairwise likelihood | 0.50394 | $5.1575 \times 10^{-4}$ |
| No Newton update | 0.50431 | $8.1279 \times 10^{-3}$ |
| M2 | 0.50018 | $1.5620 \times 10^{-2}$ |
| Wrong $\nu$ | 0.50292 | $3.5071 \times 10^{-2}$ |

REMARK. The phenomenon seen in the above simulation, that the high frequency behavior of the spectral density is important for cloud height estimation, can be interpreted in the context of Stein's work on asymptotically optimal interpolation with a mis-specified autocovariance function [see Stein (1988), Stein (1990) or Chapter 3 of Stein (1999), for example]. In this work it is noted that in many situations the low frequency behavior of the spectral density has little effect on the variance of the interpolation errors. These results are immediately applicable to cloud height estimation since constructing the super-resolution image can be viewed as an interpolation problem. In fact, this can also explain the observation noted in Section 3.1.1 that different values of $\rho$ do not severely effect the height estimates. This comes from the observation that fixing $\rho$ and estimating $\sigma$ from increments of the data (as in Sections 3.1.2 and 3.1.3) can be considered to be a data driven way to approximate the correct coefficient on the asymptotic decay of the true spectral density.

**4. Test results for cloud height retrieval.** This section shows some test results for estimating cloud height in a multi-layer cloud region. The test region is a cropped image from a MISR swath corresponding to orbit number 029145, path 031 and MISR blocks 56–67. Using the conventional grid, the upper left corner is located (row, column) = (801, 1701) and the lower right corner is located (row, column) = (1400, 2100) for a total image size of $600 \times 400$. Figure 3 shows the cloud images from the Bf and Cf cameras. This particular region was selected because of the clear nature of the two layer cloud ensemble. Near the right edge of the image, the region is dominated by optically thin, high clouds. The left edge contains mostly low, textured clouds. As one moves from left to right, the high clouds become more optically thick and the low clouds become sparse. This makes the region particularly useful for testing the ability to estimate the height of two distinct cloud layers.

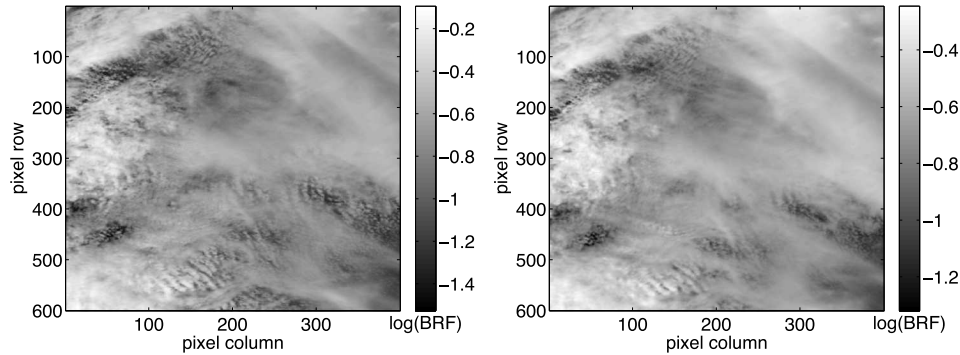

Fig. 3.   *Log BRF from the Bf camera (left) and Cf camera (right).*

In addition to the two layer structure, there is evidence to suggest minimal wind in both the along-track and across-track direction. This allows one to only estimate the height parameter and set the wind vectors to zero. In all of the test cases, the interlacing likelihood is maximized over a grid with resolution 100 meters over the interval $(0, 3 \times 10^4)$ meters. All of the following height estimates are based on 15-by-16 sliding local patches, with the exception of the 15-by-15 patch size used in the MISR software estimates and the likelihood profiles found in Figure 7. In particular, for each 15-by-16 local patch in the Bf camera, a separate likelihood is used to compute a cloud height estimate which is then associated with the midpoint of the patch. The MISR estimates use an implementation of the M2 stereo matcher which is the same algorithm used in the current MISR standard product [details can be found in Muller et al. (2002)].

Figure 4 shows the cloud-height estimates based on the low-cloud likelihood (top row) and the high-cloud likelihood (bottom row) using two cameras (left column) and three cameras (middle column). The images in the right column correspond to the differences of the height estimates from three cameras and two cameras. Figure 5 shows two of our most interesting estimates of height (utilizing different camera angles and the different likelihood techniques) along with the height estimates from MISR's M2 algorithm. The left plot in Figure 5 shows the high-cloud estimates based on the three cameras Bf, Cf and Df. The middle plot shows the low-cloud estimates based on the three cameras Aa, An and Af. The right plot shows the MISR's M2 estimates based on two cameras Bf and Cf. Compared with MISR's M2 software, one can see that the coverage (in other words, the percentage of good height estimates) is comparable. It is also apparent that the three estimates recover different components of the cloud layers. The most striking difference is the amount of high clouds recovered in the left plot compared to the other two. Presumably, a major factor in this recovery is the brightness stabilization across cameras.

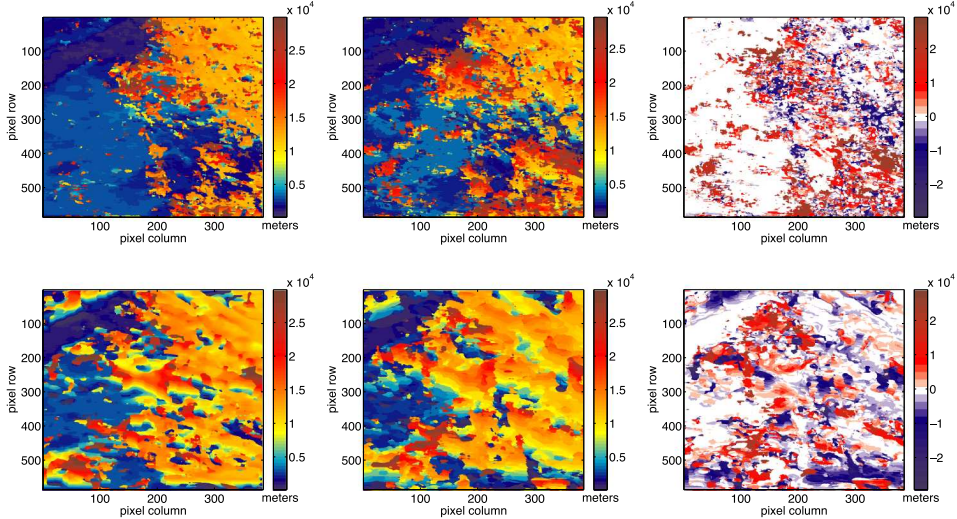

Fig. 4.   *Upper left: Low cloud-height estimates using the two cameras Bf and Cf. Upper middle: Low cloud-height estimates using the three cameras Bf, Cf and Df. Upper right: Difference of the low cloud-height estimates using two cameras versus three cameras. Lower left: High cloud-height estimates using the two cameras Bf and Cf. Lower middle: High cloud-height estimates using the three cameras Bf, Cf and Df. Lower right: Difference of the high cloud-height estimates using two cameras versus three cameras.*

Figure 6 shows the mean and variance, taken over each column of the transformed images, from cameras Bf, Cf and Df. The stabilizing transformations are found by visually matching the means and variances of the high clouds near the right edge of the test image. From these plots it seems that, at least for the test region, the high cloud region can indeed be stabilized as seen by similarity of the first two moments near the right edge, where the high clouds dominate. There is some reason to believe that such a stabilizing transformation exists for general cloud images. The radiance contribution from the high clouds is almost exclusively due to initial scattering, whereas the radiance contribution from the low clouds must pass though the high cloud region, getting a reduction in both mean and variance depending on the optical thickness of the top layer. It is not so clear, however, whether the stabilizing transformation can be estimated from a multi-layer scene. In our test case we can take advantage of the fact that high clouds dominate the right hand edge and, therefore, the overall change in mean and variance can be estimated.

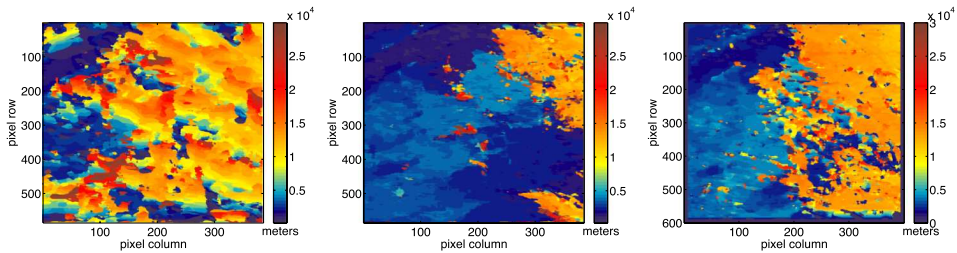

Fig. 5.   *Left: High cloud-height estimates from three cameras Bf, Cf and Df. Middle: Low cloud-height estimates from the three cameras Aa, An and Af. Right: Height estimates based on the MISR's M2 stereo matcher using cameras Bf and Cf.*

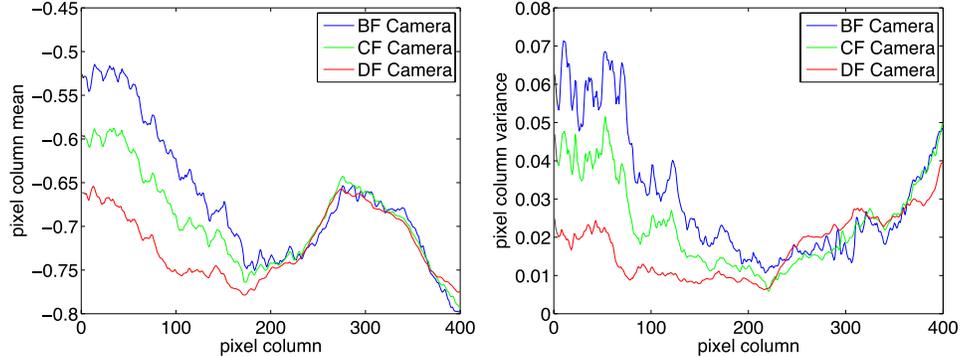

Fig. 6.  *Left: Variances of each column in the transformed images* $\mathbf{y}^{(1)}$, $1.14\mathbf{y}^{(2)} + 0.03$ *and* $1.3\mathbf{y}^{(3)} + 0.09$ *(where* $\mathbf{y}^{(1)}$, $\mathbf{y}^{(2)}$, $\mathbf{y}^{(3)}$ *denote the gray values from cameras Bf, Cf and Df respectively). Right: Mean of each column in the transformed image* $\mathbf{y}^{(1)}$, $1.14\mathbf{y}^{(2)} + 0.03$ *and* $1.3\mathbf{y}^{(3)} + 0.09$. *These stabilized gray values are used for the high-cloud estimates.*

Remark.  One possible way to automate the estimation of the stabilizing transformation is to use classification algorithms to find regions dominated mostly by high clouds, estimate the transformation on these regions, then extrapolate to the whole multi-layer scene. Indeed, there is existing literature on cloud detection which could potentially be adapted for finding these high cloud dominated regions in a multi-layer scene [see Shi et al. (2008)].

To investigate the difference between the the low-cloud likelihood and the high-cloud likelihood, Figure 7 shows the log-likelihood profiles as a function of cloud height for two patches from our test scene (both patches constitute a $15 \times 15$ pixel region). The first patch is centered at row 500 and column 100 and is dominated by low textured clouds. The second patch is centered at row 400 and column 275 and constitutes a region with a two layer cloud ensemble. In Figure 7 the top row shows plots of the low-cloud likelihood (left/right plot corresponds to the low/multi-layer cloud patch). The bottom row corresponds to the high-cloud likelihood (left/right plot corresponds to the low/multi-layer cloud patch). The discontinuous nature of the graphs is due to the fact that a fixed interlacing was used and, therefore, sub-grid-scale displacement can not be detected. The significance of these plots is that the likelihood profiles changes significantly when using the likelihood for the high-clouds (bottom row). The result is that the bottom layer height is estimated when using the low-cloud likelihood, and the top layer height is estimated when using the high-cloud likelihood. One other surprising feature of these likelihood profiles is multi-modality. We have three possible explanations for this phenomenon. First, as one searches through height space, the data from the extracted patches are different. Therefore, the likelihood is computed on different data values as one varies the cloud height parameter. Second, since the patch sizes are relatively small and the cloud images are textured, it is possible that there are multiple patches that "look the same" and hence yield a peak of the likelihood. Finally, we think that some of these peaks may correspond to multiple cloud layers in the test region.

*Summary.*   We present a new depth-of-field algorithm which uses random field models and the concept of super-resolution. By viewing the multi-angle cloud images as discrete sub-samples of a continuous random field, one can view depth-of-field estimation as a statistical parameter estimation problem. Under this paradigm, new tools become available for recovering depth-of-field from multiple stereo images and, in some cases, improve sensitivity and allow fine tuning for different observation scenarios. We apply these techniques to the recovery of cloud height using the MISR instrument on the Terra spacecraft. In our application, we attempted to demonstrate the ease in which technical details of the stereo cameras and the scientific properties of the observations can be incorporated in the estimates. The main focus of the test case was to use the special nature of our new estimator to recover the heights of a two layer cloud ensemble: an optically thin high cloud layer and an optically thick, textured low cloud. We have shown that the recovery of two separate layers is indeed possible and could potentially be automated for cloud-height estimates on a global scale. These results lay the foundation for future research on extending this framework for cloud height estimation on a global scale using all nine MISR cameras. Indeed, current research is underway to speed up the likelihood estimates and to incorporate more information on the observational properties of the MISR cameras.

**Acknowledgments.**   The research described in this paper was carried out through a collaboration between University of California, Berkeley and Jet Propulsion Laboratory, California Institute of Technology, under a contract with the National Aeronautics and Space Administration. The authors would also like to thank the Editor, the Associate Editor and the Referees for many useful comments and suggestions.

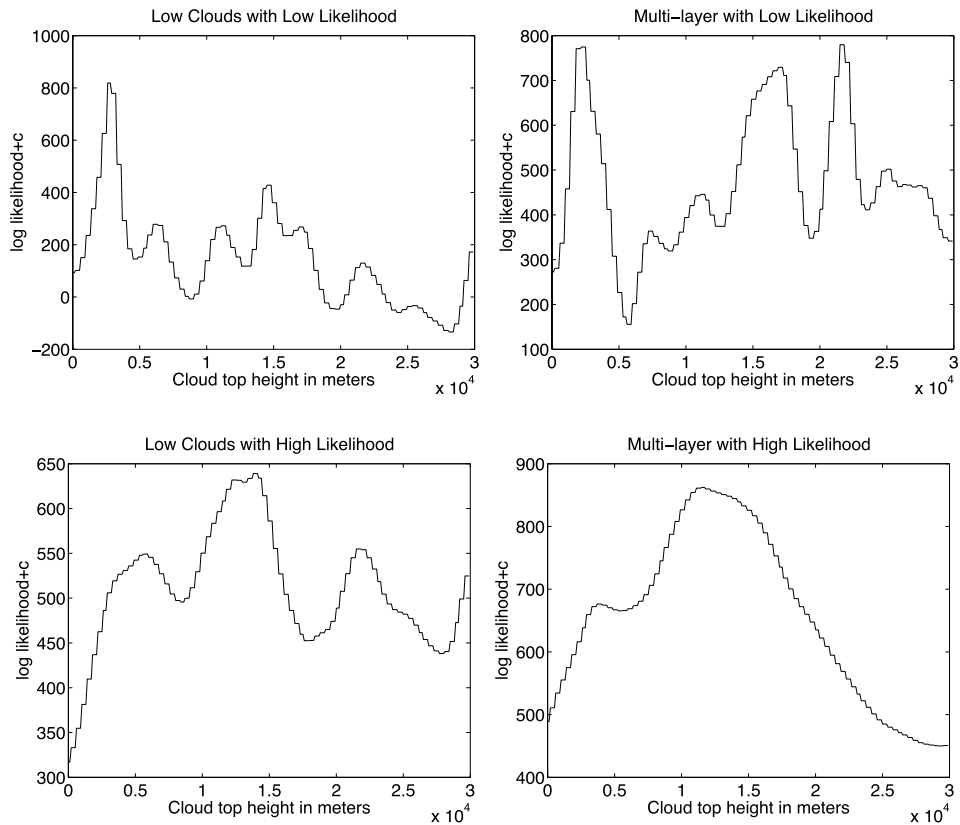

Fig. 7. *Profiles of the log likelihood as a function of height. Rows correspond to different likelihood models (top: low cloud model, bottom: high cloud model) and the columns correspond to different patches (left: low cloud patch, right: multi-layer cloud patch).*

E. ANDERES
DEPARTMENT OF STATISTICS
UNIVERSITY OF CALIFORNIA AT DAVIS
4214 MATH. SCI. BLDG.
DAVIS, CALIFORNIA 95616
USA
E-MAIL: anderes@stat.ucdavis.edu

V. JOVANOVIC
C. MORONEY
A. BRAVERMAN
JET PROPULSION LABORATORY
CALIFORNIA INSTITUTE OF TECHNOLOGY
PASADENA, CALIFORNIA 91109
USA
E-MAIL: veljko.m.jovanovic@jpl.nasa.gov
          Catherine.Moroney@jpl.nasa.gov
          Amy.Braverman@jpl.nasa.gov

B. YU
DEPARTMENT OF STATISTICS
UNIVERSITY OF CALIFORNIA AT BERKELEY
BERKELEY, CALIFORNIA 94720
USA
E-MAIL: binyu@stat.berkeley.edu

M. GARAY
INTELLIGENCE AND INFORMATION SYSTEMS
RAYTHEON CORPORATION
PASADENA, CALIFORNIA 91109
USA
E-MAIL: Michael.Garay@jpl.nasa.gov

E. CLOTHIAUX
DEPARTMENT OF METEOROLOGY
PENNSYLVANIA STATE UNIVERSITY
UNIVERSITY PARK, PENNSYLVANIA 16802
USA
E-MAIL: cloth@essc.psu.edu